\documentclass[aps,prd,preprint,groupedaddress,superscriptaddress,nofootinbib,longbibliography,showpacs,showkeys,bibnotes]{revtex4-1}
\usepackage{graphics}%
\usepackage[mathcal]{euscript}
\usepackage[latin1]{inputenc}
\usepackage{latexsym}
\usepackage{hyperref}
\usepackage{amsmath}    
\usepackage{graphicx}   
\usepackage{verbatim}  
\usepackage{color}      
\usepackage{subfigure}    
\usepackage{bm}
\usepackage{amssymb}
\usepackage{bbm}
\usepackage{float}
\usepackage{slashed}
\usepackage{amsfonts}
\usepackage{euscript}  
\usepackage{mathrsfs} 
\usepackage{bbding}
\usepackage{pifont}
\usepackage{cancel}
\usepackage{multirow}
\usepackage{soul}
\usepackage{calrsfs}
\DeclareMathAlphabet{\pazocal}{OMS}{zplm}{m}{n}

\usepackage[usenames,dvipsnames]{xcolor}
\hypersetup{colorlinks=true, citecolor=Emerald, linkcolor=Emerald,urlcolor=Emerald}

\begin{document}
%-------------------------------------------------------------------------

%-------------------------------------------------------------------------
\title{Generalized no-hair theorems without horizons}
%-------------------------------------------------------------------------

%-------------------------------------------------------------------------
%-------------------------------------------------------------------------
\author{Carlos Barcel\'o}
\email[]{carlos@iaa.es}
\address{Instituto de Astrof\'isica de Andaluc\'ia (IAA-CSIC), Glorieta de la Astronom\'ia, 18008 Granada, Spain}
\author{Ra\'ul Carballo-Rubio}
\email[]{raul.carballorubio@sissa.it}
\affiliation{SISSA - International School for Advanced Studies, Via Bonomea 265, 34136 Trieste, Italy}
\affiliation{IFPU - Institute for Fundamental Physics of the Universe, Via Beirut 2, 34014 Trieste, Italy}
\affiliation{INFN Sezione di Trieste, Via Valerio 2, 34127 Trieste, Italy} 
\author{Stefano Liberati}
\email[]{liberati@sissa.it}
\affiliation{SISSA - International School for Advanced Studies, Via Bonomea 265, 34136 Trieste, Italy}
\affiliation{IFPU - Institute for Fundamental Physics of the Universe, Via Beirut 2, 34014 Trieste, Italy}
\affiliation{INFN Sezione di Trieste, Via Valerio 2, 34127 Trieste, Italy} 

%-------------------------------------------------------------------------
%-------------------------------------------------------------------------
%-------------------------------------------------------------------------

\begin{abstract}
The simplicity of black holes, as characterized by no-hair theorems, is one of the most important mathematical results in the framework of general relativity. Are these theorems unique to black hole spacetimes, or do they also constrain the geometry around regions of spacetime with arbitrarily large (although finite) redshift? This paper presents a systematic study of this question and illustrates that no-hair theorems are not restricted to spacetimes with event horizons but are instead characteristic of spacetimes with deep enough gravitational wells, extending Israel's theorem to static spacetimes without event horizons that contain small deviations from spherical symmetry. Instead of a uniqueness result, we obtain a theorem that constrains the allowed deviations from the Schwarzschild metric and guarantees that these deviations decrease with the maximum redshift of the gravitational well in the external vacuum region. Israel's theorem is recovered continuously in the limit of infinite redshift. This result provides a first extension of no-hair theorems to ultracompact stars, wormholes, and other exotic objects, and paves the way for the construction of similar results for stationary spacetimes describing rotating objects.
\end{abstract}

%-------------------------------------------------------------------------
\maketitle
%-------------------------------------------------------------------------

%------------------------------------------------------------------------------------------------
\section{Introduction}
%------------------------------------------------------------------------------------------------

No-hair theorems \cite{Israel1967,Carter1971,Robinson1975} are probably one of the best-known mathematical results in the theory of general relativity, for both specialists and non-specialists alike. The suggestive metaphor coined by Wheeler has been used during decades in order to motivate further research and capture the imagination of the general public. There are still many aspects to be understood regarding these theorems, from mathematical technicalities \cite{Wald1984,Carter1987,Carter1997,Mazur2000} to the study of their interplay with observations \cite{Johannsen2010,Bambi2015,Cardoso2016,Krishnendu2017,Thrane2017}.

We deal here with new aspects that extend the applicability of these results. No-hair theorems have been associated with event horizons (infinite-redshift surfaces) in classical general relativity, restricting the number of independent multipole moments that are needed in order to characterize black holes. In static situations, to which we limit our present discussion, Israel's theorem \cite{Israel1967} establishes that the event horizon must be spherically symmetric: a static and isolated black hole cannot be deformed away from sphericity.

The very idea of testing observationally no-hair theorems pushes to the limit the essence of experimental confirmation. Experimentally, the best one would be able to do is to place constraints on the size of deformations around astrophysical black holes, and correlate these constraints with other observables such as the gravitational redshift or the spacetime curvature. Ideally, one would like to be able to carry out these tests without the need of assuming the existence of event horizons, given that these tests will be carried out on finite-redshift surfaces.

However, no-hair theorems do not apply unless the existence of an event horizon is assumed. This implies that the multipolar structure of spacetimes that do not contain an infinite-redshift surface, but only surfaces in which the redshift is extremely large although finite (these can share most of the observational properties of actual black holes \cite{Cardoso2017,Carballo-Rubio2018b} and are therefore often called ``black hole mimickers''), is not constrained by no-hair theorems and may therefore be arbitrary in principle. This may seem to open the possibility of having largely different multipolar structures, that would point (if detected) to a non-Kerr nature of astrophysical black holes; it is worth mentioning that experiments such as the Event Horizon Telescope are starting to probe regions in which tests of this kind would be possible \cite{Giddings2016}. However, in this work we show that the existence of surfaces with extremely large redshift places by itself strong limits on the allowed multipolar structures. This has implications for the search of physics beyond general relativity, but also for the understanding of the physics behind no-hair theorems.

As our goal in this paper is generalizing Israel's theorem \cite{Israel1967} to spacetimes without event horizons, we have decided to follow as close as possible the notation and conventions in this seminal work, in order to facilitate comparisons and stress the novelty of the present discussion. This is also important given that Israel's theorem is the starting point for more general no-hair theorems dealing with charged and rotating black holes; sticking to the original notation may also simplify extending our results to these situations.

%------------------------------------------------------------------------------------------------
\section{Setting}
%------------------------------------------------------------------------------------------------

We will be working with static spacetimes, characterized by a hypersurface-orthogonal timelike Killing vector field $\xi$. The line element can be written as
\begin{equation}
\text{d}s^2=-V^2(x^1,x^2,x^3)\text{d}t^2+g_{\alpha\beta}(x^1,x^2,x^3)\text{d}x^\alpha\text{d}x^\beta,\label{eq:4dmetric}
\end{equation}
where $V^2=|\xi|^2>0$ and the coordinate $t$ has been chosen such that $\nabla t=\xi V^{-2}$. In the following, Greek indices run from 1 to 3, lower case italic indices from 1 to 2, and upper case italic indices from 0 to 3.

The hypersurfaces $\Sigma$ are defined by constant values of $t$. The induced metric is given by
\begin{equation}
\left.\text{d}s^2\right|_{\Sigma}=g_{\alpha\beta}(x^1,x^2,x^3)\text{d}x^\alpha\text{d}x^\beta.\label{eq:3dmetric}
\end{equation}
We now choose $V$ as one of the coordinates in the hypersurfaces $\Sigma$, so that in the following we will write $(x^1,x^2,x^3)=(V,\theta^1,\theta^2)$. Moreover, we can choose the coordinates $(\theta^1,\theta^2)$ on $\mathscr{B}_V$ to be orthogonal to the equipotential surfaces, $g^{\alpha\beta}\partial_\alpha\theta^a\partial_\beta V=0$. The (spacelike) normal vector to the two-dimensional subspaces $\mathscr{B}_V$ of $\Sigma$ defined by constant values of $V$ is given by
\begin{equation}
n_\alpha=\rho\partial_\alpha V,
\end{equation}
where
\begin{equation}
\rho=\frac{1}{\sqrt{g^{\alpha\beta}\partial_\alpha V\partial_\beta V}}.
\end{equation}
This function $\rho$ is related to the trace of the extrinsic curvature $K$ of the two-dimensional subspaces $\mathscr{B}_V$ as
\begin{equation}
\frac{\partial\rho}{\partial V}=\rho^2 K.\label{eq:rel4}
\end{equation}
The metric in Eq. \eqref{eq:3dmetric} can be written as \cite{Israel1967,Israel1967b}
\begin{equation}
\left.\text{d}s^2\right|_{\Sigma}=g_{ab}(V,\theta^1,\theta^2)\text{d}\theta^a\text{d}\theta^b+\rho^2(V,\theta^1,\theta^2)\text{d}V^2.\label{eq:3dmetricb}
\end{equation}
%

%------------------------------
\begin{figure}[h]%
\caption{Two-dimensional subspaces in $\Sigma$ defined by constant values of $V$. The internal region has been emptied in order to emphasize that the proof below only needs to consider the vacuum spacetime region with $V\geq\epsilon$. The white region will generally have a nonzero stress-energy tensor that violates the classical energy conditions, being filled with horizonless objects such as wormholes, gravastars or black stars.}\label{fig:fig1}
\vspace{0.4cm}
\includegraphics[width=0.4\textwidth]{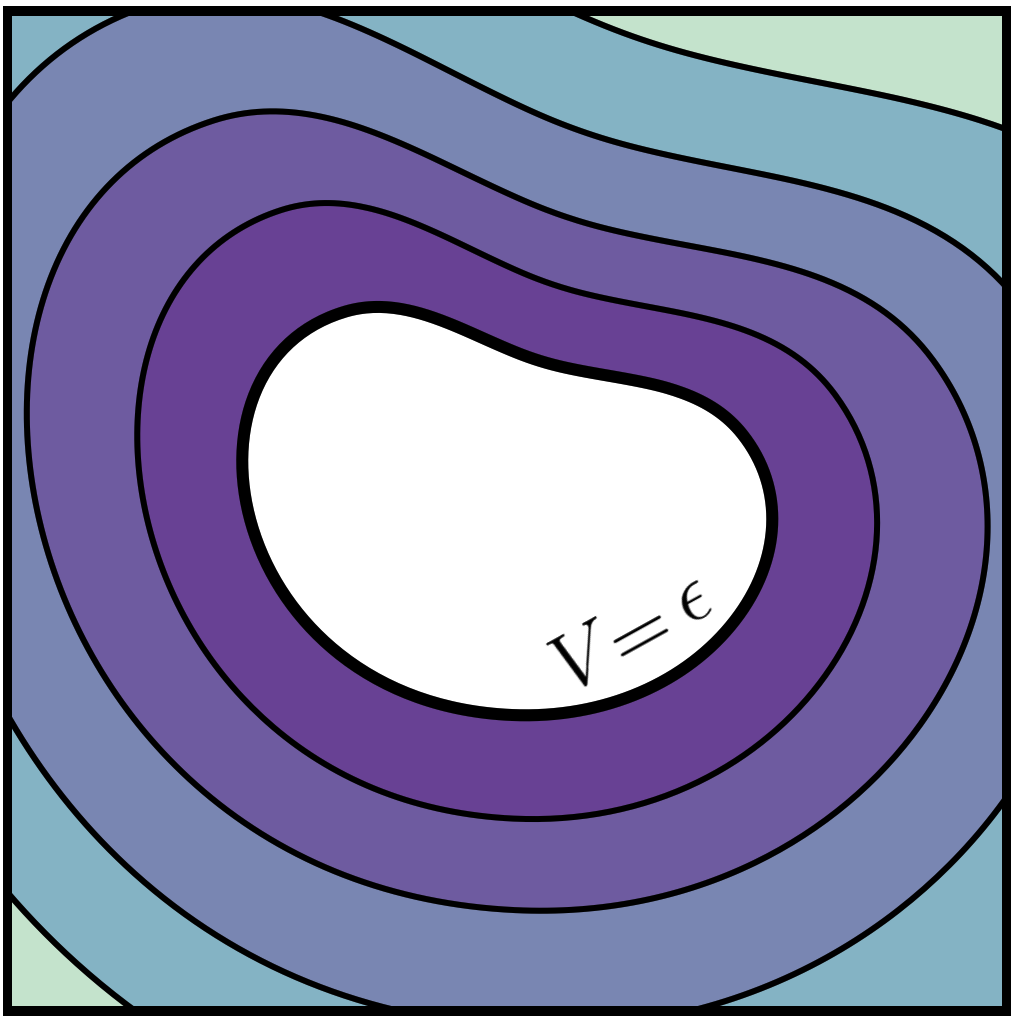}
\end{figure}%
%------------------------------

\noindent
In this setting, there are several relations of purely geometric origin (but that hold only for static vacuum spacetimes) that were derived in \cite{Israel1967}. For the sake of conciseness we shall not duplicate their derivation here. In particular, there are three relations that include a total derivative with respect to $V$ of the functions introduced above, which will be useful for our discussion. The first of such relations is
\begin{equation}
\frac{\partial}{\partial V}\left(\frac{g^{1/2}}{\rho}\right)=0,\label{eq:rel1}
\end{equation}
where we have used the notation $g=\mbox{det}(g_{ab})$ (to which we will stick troughout the paper). The second and third ones are, respectively,
\begin{equation}
\frac{\partial}{\partial V}\left(\frac{g^{1/2}}{\rho^{1/2}}\frac{K}{V}\right)=-\frac{2g^{1/2}}{V}\left[g^{ab}\nabla_a\nabla_b(\rho^{1/2})+\frac{1}{\rho^{3/2}}\left(\frac{1}{2}\partial_a\rho\partial^a\rho+\psi_{ab}\psi^{ab}\right)\right]\label{eq:rel2}
\end{equation}
and
\begin{equation}
\frac{\partial}{\partial V}\left[\frac{g^{1/2}}{\rho}\left(KV+\frac{4}{\rho}\right)\right]+g^{1/2}Vg^{ab}R_{ab}=-g^{1/2}V\left[g^{ab}\nabla_a\nabla_b(\ln\rho)+\frac{1}{\rho^2}\left(\partial_a\rho\partial^a\rho+2\psi_{ab}\psi^{ab}\right)\right].\label{eq:rel3}
\end{equation}
In the two last equations we have defined
\begin{equation}
\psi_{ab}=\rho\left(K_{ab}-\frac{1}{2}g_{ab}K\right).\label{eq:psidef}
\end{equation}
It will be useful later to note that $\psi_{ab}=0$ and $\partial_a\rho=0$ imply spherical symmetry \cite{Israel1967}. The three relations \eqref{eq:rel1}, \eqref{eq:rel2} and \eqref{eq:rel3} are valid for any vacuum static spacetime. We will eventually integrate these equations on $\Sigma$. It will be then useful to recall that
\begin{equation}\label{eq:top0}
\int\text{d}\theta^1\text{d}\theta^2\,g^{1/2}g^{ab}\nabla_a\nabla_b f=0,
\end{equation}
where $g^{ab}\nabla_a\nabla_b$ is the two-dimensional Laplace operator on the two-dimensional subspace spanned by $(\theta^1,\theta^2)$, and $f=f(\theta^1,\theta^2)$ an arbitrary function of these variables. Also, the Gauss-Bonnet theorem implies (note that the Gaussian curvature is $\mathcal{K}=g^{ab}R_{ab}/2$) that
\begin{equation}\label{eq:top1}
\int\text{d}\theta^1\text{d}\theta^2\,g^{1/2}g^{ab}R_{ab}=-8\pi.
\end{equation}
The last geometric relation that will be needed below is the decomposition of the (four-dimensional) Kretschmann scalar as
\begin{equation}
R_{ABCD}R^{ABCD}=\frac{8}{V^2\rho^2}\left[\frac{2}{\rho^2}\partial_a\rho\partial^a\rho+K_{ab}K^{ab}+\frac{1}{\rho^4}\left(\frac{\partial\rho}{\partial V}\right)^2\right].\label{eq:krets}
\end{equation}
Let us now formulate the main result in this paper.

%------------------------------------------------------------------------------------------------
\section{Statement}\label{sec:statement}
%------------------------------------------------------------------------------------------------

Let us make the following assumptions (the first two ones are exactly the same as in \cite{Israel1967}):
\begin{itemize}
\item[(a)]{The spacetime is static, foliated by spatial hypersurfaces from which we can take any representative $\Sigma$.}
\item[(b)]{$\Sigma$ is regular, empty, noncompact, and asymptotically Euclidean, in the following sense: there exists a coordinate system in which the metric \eqref{eq:4dmetric} has the asymptotic form
\begin{align}
g_{\alpha\beta}=\delta_{\alpha\beta}+\mathscr{O}(r^{-1}),\qquad \partial_\gamma g_{\alpha\beta}=\mathscr{O}(r^{-2}),\qquad V=1-\frac{m}{r}+\eta,
\end{align}
where $m\in\mathbb{R}$, $\eta=\mathscr{O}(r^{-2})$, $\partial_\alpha\eta=\mathscr{O}(r^{-3})$, $\partial_\alpha\partial_\beta\eta=\mathscr{O}(r^{-4})$ and $r=(\delta_{\alpha\beta}x^\alpha x^\beta)\rightarrow\infty$ or, equivalently, $V\rightarrow1$}.
\item[(c)]{The equipotential surfaces with constant $V$ in $\Sigma$, $\mathscr{B}_V$, are regular, simply connected and closed two-dimensional spaces with finite area $\mathscr{S}_V$, the shapes of which deviate slightly from spherical symmetry.}
\item[(d)]{The spacetime is vacuum for $V\geq\epsilon$, with $0<\epsilon\ll 1$ parametrizing the redshift of the innermost equipotential surface that is strictly outside the non-vacuum internal region (in the case of a black hole the relevant innermost equipotential surface corresponds to the event horizon, which means that $\epsilon=0$ for a black hole).}
\item[(e)]{The Kretschmann scalar $R_{ABCD}R^{ABCD}$ is bounded from above, for instance due to the Einstein field equations being valid only up to a certain critical value of the curvature.}
\end{itemize}
As we demonstrate below, under these conditions one can show that the vacuum spacetime region is arbitrarily close to the Schwarzschild solution in the $\epsilon\rightarrow 0$ limit. This extends Israel's theorem, in the sense that the latter is naturally included as the particular case $\epsilon=0$.

%------------------------------------------------------------------------------------------------
\section{Proof}
%------------------------------------------------------------------------------------------------

%------------------------------------------------------------------------------------------------
\subsection{Preliminaries}
%------------------------------------------------------------------------------------------------

Let us start with a couple of considerations regarding the coordinates $(\theta^1,\theta^2)$. The form of $g_{ab}(V,\theta^1,\theta^2)$ in Eq. \eqref{eq:3dmetricb} can be further constrained using the remaining freedom in the choice of these coordinates. From our assumptions above, we can deduce that it is always possible to choose coordinates $(\hat{\theta}^1,\hat{\theta}^2)$ such that they correspond to the usual spherical coordinates in the limit $r\rightarrow\infty$ (or $V\rightarrow1$). Eq. \eqref{eq:rel1} implies then that $g^{1/2}$ and $\rho$ are related by a multiplicative function that is independent of $V$, which is univocally determined by the asymptotic conditions (b) so that we can write
\begin{equation}
\sqrt{g(V,\hat{\theta}^1,\hat{\theta}^2)}=m\rho(V,\hat{\theta}^1,\hat{\theta}^2)\sin\hat{\theta}^1.\label{eq:rel5}
\end{equation}
The normalization constant $m$ is determined from the asymptotic behavior of the two functions of $V$ involved. The previous equation suggests that it would be useful to introduce a normalized metric $h_{ab}$ as
\begin{equation}\label{eq:metrel}
g_{ab}(V,\hat{\theta}^1,\hat{\theta}^2)=m\rho(V,\hat{\theta}^1,\hat{\theta}^2)h_{ab}(V,\hat{\theta}^1,\hat{\theta}^2),
\end{equation}
which satisfies the constraint $\mbox{det}(h_{ab})=\sin\hat{\theta}^1$. Eq. \eqref{eq:3dmetricb} reads then
\begin{equation}
\left.\text{d}s^2\right|_{\Sigma}=m\rho(V,\hat{\theta}^1,\hat{\theta}^2)h_{ab}\text{d}\hat{\theta}^a\text{d}\hat{\theta}^b+\rho^2(V,\hat{\theta}^1,\hat{\theta}^2)\text{d}V^2.\label{eq:3dmetricc}
\end{equation}
%

%------------------------------------------------------------------------------------------------
\subsection{Spacetimes that are almost spherically symmetric}\label{sec:curbound}
%------------------------------------------------------------------------------------------------

From Israel's theorem \cite{Israel1967}, which is the particular case $\epsilon=0$ of the statement in Sec. \ref{sec:statement}, we know that for $\epsilon=0$ the coordinates $(\hat{\theta}^1,\hat{\theta}^2)$ are the usual spherical coordinates on the 2-sphere. Also, spherical symmetry demands that $\rho$ is not a function of these angular coordinates, with the Schwarzschild solution arising for a specific functional relation $\rho(V)$. 

In this paper, we want to understand the behaviour of spacetime when small deviations from spherical symmetry are introduced. From Israel's theorem, it is intuitively natural to expect that these perturbations must be proportional to $\epsilon$. How this intuition is realized in a precise way, and its connection to condition (e) in Sec. \ref{sec:statement}, will be discussed in the following sections.

In this section, we introduce a dimensionless perturbation parameter $X$ that characterizes the deviations from spherical symmetry, and study the behavior of the relevant functions, namely $\rho(V,\hat{\theta}^1,\hat{\theta}^2)$ and $h_{ab}(V,\hat{\theta}^1,\hat{\theta}^2)$, in terms of this parameter. The metric $h_{ab}$ will be simply given by
\begin{equation}\label{eq:pertmet}
h_{11}=1+\mathscr{O}(X),\qquad h_{22}=\sin^2\hat{\theta}^1+\mathscr{O}(X),\qquad h_{12}=h_{21}=\mathscr{O}(X).
\end{equation}
The subleading terms $\mathscr{O}(X)$ in the equation above cannot be independent in order to satisfy the constraint $\mbox{det}(h_{ab})=\sin\hat{\theta}^1$ (that must be satisfied at all orders in $X$), but it is not necessary to derive the corresponding relations for our discussion. Let us stress that, in general, $X=X(V)$ is a function of the coordinate $V$, although we do not write this dependence explicitly in order to simplify the notation in the equations below.

On the other hand, we can write
\begin{equation}\label{eq:pertrho}
\rho(V,\hat{\theta}^1,\hat{\theta^2})=\rho(V)+\rho(V)\sum_{l=0}^\infty\sum_{m=-l}^l\sigma_{lm}(V)Y_{lm}(\hat{\theta}^1,\hat{\theta^2}),
\end{equation}
where $Y_{lm}(\hat{\theta}^1,\hat{\theta^2})$ are the usual spherical harmonics and $\sigma_{lm}(V)=\mathscr{O}(X)$ are dimensionless coefficients due to the introduction of $\rho(V)$ in front of the summation symbols.

Also, exploiting the transformation of the Ricci scalar under a conformal transformation, it follows that (note the sign convention for the Ricci tensor $R_{ab}=R^c_{\ abc}$)
\begin{equation}
g^{ab}R_{ab}(V,\hat{\theta}^1,\hat{\theta^2})=-\frac{2}{m\rho(V,\hat{\theta}^1,\hat{\theta^2})}+\mathscr{O}(X).\label{eq:ricsca}
\end{equation}
A last useful relation for the trace of the extrinsic curvature $K$ can be obtained using the previous equation as well as the geometric identity
\begin{equation}\label{eq:rel6}
g^{ab}R_{ab}=K_{ab}K^{ab}-K^2-\frac{2}{\rho}\frac{K}{V}.
\end{equation}
First of all, in the presence of deviations from spherical symmetry we will have
\begin{equation}
K_{ab}=\frac{1}{2}Kg_{ab}+\mathscr{O}(X).
\end{equation}
Hence, Eq. \eqref{eq:rel6} becomes
\begin{equation}\label{eq:rel6b}
g^{ab}R_{ab}=-\frac{1}{2}K^2-\frac{2}{\rho}\frac{K}{V}+\mathscr{O}(X).
\end{equation}
Evaluating this equation on $V=\epsilon\ll1$ and taking into account Eq. \eqref{eq:ricsca}, we can see that the first term on the right-hand side of Eq. \eqref{eq:rel6b} is $\mathscr{O}(\epsilon)$ and therefore subdominant, so that we can write
\begin{equation}\label{eq:tracek}
\frac{K(\epsilon,\hat{\theta}^1,\hat{\theta}^2)}{\epsilon}=-\frac{1}{2}\rho(\epsilon,\hat{\theta}^1,\hat{\theta}^2)\left.g^{ab}R_{ab}\right|_{V=\epsilon}+\mathscr{O}(X_\epsilon,\epsilon^2)=\frac{1}{m}+\mathscr{O}(X_\epsilon,\epsilon^2).
\end{equation}
In this equation, we have defined $X_\epsilon=X(\epsilon)$. We will keep using this notation in the equations below, also for other functions of $V$.

%------------------------------------------------------------------------------------------------
\subsection{Imposing the boundary conditions}
%------------------------------------------------------------------------------------------------

We can now use the relations derived in Sec. \ref{sec:curbound} in order to integrate the three identities \eqref{eq:rel1}, \eqref{eq:rel2} and \eqref{eq:rel3}:
\begin{itemize}
\item{Let us start with Eq. \eqref{eq:rel1}, from which (together with the asymptotic conditions in our main statement) we obtained Eq. \eqref{eq:rel5}. Integrating this equation on the 2-sphere we obtain
\begin{equation}
\mathscr{S}_V=\int_{\mathscr{B}_V}\text{d}\hat{\theta}^1\text{d}\hat{\theta}^2\sqrt{g}=4\pi m\langle\rho(V,\hat{\theta}^1,\hat{\theta}^2)\rangle,
\end{equation}
where the average on the right-hand side is the standard one on the 2-sphere. This general equation reduces, for perturbative deviations with respect to spherical symmetry, to
\begin{equation}\label{eq:res1}
\mathscr{S}_V=4\pi m\rho(V)+\mathscr{O}(X).
\end{equation}
This is the relation that one would expect, on the basis of Eq. \eqref{eq:3dmetricc}, in situations close to spherical symmetry.}

\item{Let us now turn our attention to Eq. \eqref{eq:rel2} and integrate it on the interval $V\in[\epsilon,1]$ as well as on the angular variables,
\begin{equation}\label{eq:firstj}
\int_\epsilon^1\text{d}V\int_{\mathscr{B}_V}\text{d}\hat{\theta}^1\text{d}\hat{\theta}^2\frac{\partial}{\partial V}\left(\frac{g^{1/2}}{\rho^{1/2}}\frac{K}{V}\right)=8\pi\sqrt{m}-\int_{\mathscr{B}_V}\text{d}\hat{\theta}^1\text{d}\hat{\theta}^2\left.\frac{g^{1/2}}{\rho^{1/2}}\frac{K}{V}\right|_{V=\epsilon}=-P(\epsilon)\leq0,
\end{equation}
where the asymptotic value ($V=1$) has been evaluated using the asymptotic form of the spacetime metric imposed in Sec. \ref{sec:statement}, and we have used Eq. \eqref{eq:top0} in order to simplify the integral of the right-hand side of Eq. \eqref{eq:rel2} (the equation being integrated) and define
\begin{equation}\label{eq:jdef}
P(\epsilon)=2\int_\epsilon^1\text{d}V\int_{\mathscr{B}_V}\text{d}\hat{\theta}^1\text{d}\hat{\theta}^2\,\frac{g^{1/2}}{V\rho^{3/2}}\left(\frac{1}{2}\partial_a\rho\partial^a\rho+\psi_{ab}\psi^{ab}\right)\geq0.
\end{equation}
Direct substitution of Eqs. \eqref{eq:rel5} and \eqref{eq:tracek} leads to
\begin{equation}
8\pi\sqrt{m}\leq4\pi\sqrt{\rho_\epsilon}+\mathscr{O}(X_\epsilon,\epsilon).\label{eq:res2}
\end{equation}
}
\item{The same integration over Eq. \eqref{eq:rel3} allows us, using Eq. \eqref{eq:top1}, to write
\begin{equation}
-\int_{\mathscr{B}_V}\text{d}\hat{\theta}^1\text{d}\hat{\theta}^2\left.\frac{g^{1/2}}{\rho}\left(KV+\frac{4}{\rho}\right)\right|_{V=\epsilon}+4\pi=-Q(\epsilon)\leq0,\label{eq:rel3b}
\end{equation}
where we have taken into account Eq. \eqref{eq:top0} in order to simplify the integral of the right-hand side of Eq. \eqref{eq:rel3}, which can be written simply as
\begin{equation}
Q(\epsilon)=\int_\epsilon^1\text{d}V\int_{\mathscr{B}_V}\text{d}\hat{\theta}^1\text{d}\hat{\theta}^2\frac{g^{1/2}V}{\rho^2}\left(\partial_a\rho\partial^a\rho+2\psi_{ab}\psi^{ab}\right)\geq0.
\end{equation}
We just now need to use Eq. \eqref{eq:rel5} and take into account that the $KV$ term is subleading, in order to write
\begin{equation}
\rho_\epsilon\leq 4m+\mathscr{O}(X_\epsilon).\label{eq:res3}
\end{equation}
}
\end{itemize}
\noindent
We can combine these relations in order to obtain stronger statements. For instance, Eqs. \eqref{eq:res2} and \eqref{eq:res3} imply that
\begin{equation}\label{eq:rho4m}
\rho_\epsilon=4m+\mathscr{O}(X_\epsilon,\epsilon),
\end{equation}
which, in turn, makes Eq. \eqref{eq:res1} equivalent to
\begin{equation}
\mathscr{S}_\epsilon=16\pi m^2+\mathscr{O}(X_\epsilon,\epsilon).\label{eq:areares}
\end{equation}
Moreover, we can use the fact that the integrands in the definitions of both $P(\epsilon)$ and $Q(\epsilon)$ are definite positive to extract constraints on $\partial_a\rho$ and $\psi_{ab}$. In Israel's theorem both $P(\epsilon)$ and $Q(\epsilon)$ have to vanish due to the corresponding version of Eq. \eqref{eq:rho4m}. For the present discussion, it is only $P(\epsilon)$ which is useful, due to the different dependence on $V$ of both integrands. This is, however, enough to derive the constraints
\begin{equation}\label{eq:rhopsiconst}
\partial_a\rho\partial^a\rho=\mathscr{O}(X_\epsilon,\epsilon),\qquad \psi_{ab}\psi^{ab}=\mathscr{O}(X_\epsilon,\epsilon).
\end{equation}
This concludes the proof that any spacetime satisfying the assumptions previously stated is a perturbation of the Schwarzschild solution.

%------------------------------------------------------------------------------------------------
\section{Calculating the strength of multipoles \label{eq:multi}}
%------------------------------------------------------------------------------------------------

In order to understand the physical implications of our result above, it is useful to characterize in terms of multipoles the strength of the deviations from spherical symmetry proportional to the parameter $X$. The vacuum Einstein field equations imply that the function $V$ in Eq. \eqref{eq:4dmetric} satisfies \cite{Israel1967}
\begin{equation}\label{eq:lapeq0}
g^{\alpha\beta}\nabla_\alpha\nabla_\beta V=0.
\end{equation}
The relevant equation to solve is the Laplace equation in the Schwarzschild background for $l=0$,
\begin{equation}
\sqrt{1-\frac{2m}{r}}\frac{\text{d}}{\text{d} r}\left(r^2\sqrt{1-\frac{2m}{r}}\frac{\text{d} V}{\text{d} r}\right)=0.
\end{equation}
It is straightforward to show that its solution is given by
\begin{equation}
V=V(r)=\sqrt{1-\frac{2m}{r}}.
\end{equation}
We want to understand the behavior of perturbations with respect to this spherically symmetric solution with angular variables $(\theta,\phi)$,
\begin{equation}
V(r,\theta,\phi)=V(r)+\sum_{l=1}^\infty\sum_{m=-l}^lf_{lm}(r)Y_{lm}(\theta,\phi),
\end{equation}
which, in a perturbative treatment, verify
\begin{equation}\label{eq:pertdiffeq}
\sqrt{1-\frac{2m}{r}}\frac{\text{d}}{\text{d} r}\left(r^2\sqrt{1-\frac{2m}{r}}\frac{\text{d} f_{lm}}{\text{d} r}\right)-l(l+1)f_{lm}=0.
\end{equation}
Solving this equation allows us to understand the asymptotic behavior of multipole moments that are perturbative at $V=\epsilon$ (and, therefore, remain perturbative when the redshift decreases). Let us perform some algebraic manipulations in order to find the general solution to this equation, starting with the change of variables
\begin{equation}
z=\frac{r}{m}-1\geq 1,
\end{equation}
which allows us to rewrite Eq. \eqref{eq:pertdiffeq} as
\begin{equation}
\frac{\text{d}}{\text{d}z}\left[(z^2-1)\frac{\text{d}f_{lm}}{\text{d}z}\right]-\frac{\text{d}f_{lm}}{\text{d}z}-l(l+1)f_{lm}=0.
\end{equation}
Now we can redefine
\begin{equation}
f_{lm}(z)=\left(\frac{z-1}{z+1}\right)^{1/4}y_{lm}(z),
\end{equation}
which results in
\begin{equation}
\frac{\text{d}}{\text{d}z}\left[(z^2-1)\frac{\text{d}y_{lm}}{\text{d}z}\right]-l(l+1)y_{lm}-\frac{1}{4}\frac{1}{z^2-1}y_{lm}=0.
\end{equation}
This equation is a particular case of the associated (or general) Legendre differential equation \cite{Olver2010} (the standard Legendre equation would be obtained dropping the last term). In the interval of interest, namely $z\in(1,\infty)$, there are two independent solutions $P^{1/2}_l(z)$ and $Q^{1/2}_l(z)$ \cite{Olver2010}. However, the condition of asymptotic flatness permits us to discard the first one, which is divergent in the $z\rightarrow\infty$ (i.e., $r\rightarrow\infty$) limit. Hence, the relevant solution to the previous equation is
\begin{equation}
y_{lm}(z)=C_lQ^{1/2}_l(z),
\end{equation}
where $C_l$ is an integration constant. Hence,
\begin{equation}
f_{lm}(r)=C_l\left(1-\frac{2m}{r}\right)^{1/4}Q^{1/2}_{l}(r/m-1).
\end{equation}
All we need is the asymptotic behavior of this special function, which is given by \cite{Olver2010}:
\begin{equation}\label{eq:asymp1+}
Q^{1/2}_l(z)\simeq\frac{\sqrt{\pi}}{2\Gamma(l+3/2)}\left(\frac{z+1}{z-1}\right)^{1/4},\qquad z\rightarrow1^+\qquad (r\rightarrow 2m),
\end{equation}
and
\begin{equation}\label{eq:asympinf}
Q^{1/2}_l(z)\simeq\frac{\sqrt{\pi}}{\Gamma(l+3/2)}\frac{1}{(2z)^{l+1}},\qquad z\rightarrow\infty\qquad (r\rightarrow\infty).
\end{equation}
We now want to impose that the value of $V(r,\theta,\phi)$ at $r=r_\epsilon=2m/(1-\epsilon^2)$ represents a perturbative deviation with respect to the Schwarzschild geometry, namely
\begin{equation}
V(r_\epsilon,\theta,\phi)=\epsilon\left[1+\mathscr{O}(X_\epsilon)\right].
\end{equation}
From this imposition and Eq. \eqref{eq:asymp1+}, and taking into account that the different $(1-z)$ factors in $f_{lm}(r)$ cancel in the $z\rightarrow1^+$ limit, it follows that
\begin{equation}
\lim_{r\rightarrow 2m}f_{lm}(r)\simeq \frac{\sqrt{\pi}}{2\Gamma(l+3/2)} C_l,
\end{equation}
so that the constant of integration $C_l$ must verify
\begin{equation}\label{eq:mults}
C_l\propto \epsilon X_\epsilon+\mathscr{O}(X_\epsilon^2).
\end{equation}
On the other hand, we can see explicitly from Eq. \eqref{eq:asympinf} that this constant of integration $C_l$ gives precisely the (dimensionless) strength of the mass multipoles $M_l$ in the asymptotically flat region of spacetime (e.g., \cite{Cardoso2016}). It is straightforward to see that, for $X\ll1$, only perturbative multipoles can be induced (which is certainly reasonable).

%------------------------------------------------------------------------------------------------
\section{Implications and conclusions \label{sec:imp}}
%------------------------------------------------------------------------------------------------

We are now in position of extracting the physics of the results above. In qualitative terms, one just needs three independent parameters to describe the situation: $m$, the mass measured asymptotically; $\epsilon$, the parameter measuring the compactness of the central object; and $X$, the parameter that controls the deviations from spherical symmetry (alternatively, we can use the strength of the mass multipoles $M_l$). The discussion above relies on a perturbative expansion in the parameter $X$ which, when valid, implies that deviations with respect to sphericity are small enough\footnote{Let us stress that this is indeed the kind of physical situation one is most interested in; moreover, it is this situation which illustrates more sharply that even tiny deviations from spherical symmetry have dramatic consequences for compact enough configurations.}.

The main conclusion that we want to highlight here is that even these small perturbations from sphericity can have significant effects on the curvature. In fact, plugging Eqs. \eqref{eq:pertrho} and \eqref{eq:rho4m} into the first term on the right-hand side of Eq. \eqref{eq:krets}, it is straightforward to see that
\begin{equation}\label{eq:mc1}
\left.R_{ABCD}R^{ABCD}\right|_{V=\epsilon}\gtrsim\frac{1}{m^4}\left(\frac{X_\epsilon^2}{\epsilon^2}\right).
\end{equation}
The first multiplicative factor on the right-hand side of this equation is the typical value of the curvature in the surroundings of the gravitational radius of a black hole with mass $m$. The second factor depends on the depth of the gravitational well $\epsilon$ and the shape of the closed surface $V=\epsilon$. If this closed surface is spherically symmetric ($X_\epsilon=0$), no additional contributions to the curvature are generated. However, if one deforms continuously this closed surface up to $\epsilon\ll X_\epsilon\ll 1$, the curvature largely grows.

Let us introduce a scale $L$ that determines the typical curvature generated for fixed values of the two parameters $X_\epsilon$ and $\epsilon$. Also, we can use Eq. \eqref{eq:mults} to determine that the typical strength of the corresponding multipoles is $M_l\propto \epsilon X_\epsilon$. It follows that
\begin{equation}\label{eq:mc2}
\frac{1}{L^2}\gtrsim\frac{1}{m^2}\sum_{l=1}^\infty\left(\frac{M_l}{\epsilon^2}\right).
\end{equation}
We can illustrate the meaning of these two equations \eqref{eq:mc1} and \eqref{eq:mc2} by considering an initial condition given by a spherically symmetric static star with $m=\mathscr{O}(M_\odot)$ and a given compactness (as measured by $\epsilon$), and discussing what happens when applying a deformation:
\begin{itemize}
\item[1)]{$\epsilon=\mathscr{O}(1)$: this would correspond to the compactness of a typical neutron star. It is possible to modify the shape of these configurations and introduce multipoles of $\mathscr{O}(1)$ without changing the order of magnitude of the curvature, given by $1/m^2$.}
\item[2)]{$\epsilon=\ell/m\ll1$: this corresponds to a structure that is a Planck length $\ell$ away (as measured in the proper radial length) from forming a horizon (there are different proposals of structures with this compactness; e.g., \cite{Mazur2004,Mottola2010,Visser2003,Barcelo2007,Barcelo2009,Carballo-Rubio2017}). This structure becomes more stiff, in the sense that even introducing tiny multipoles proportional to $\epsilon$ induces curvatures of order
\begin{equation}
\frac{1}{m^2}\left(\frac{m}{\ell}\right)\gg \frac{1}{m^2}.
\end{equation}
While the above observation holds on the basis of our perturbative analysis in $X_\epsilon$, it is interesting to note that there is no reason to expect that the validity of Eqs. \eqref{eq:mc1} and \eqref{eq:mc2} could not be extended to situations in which $X_\epsilon$ takes greater values such that $M_l\sim 1$. One should conclude then that $\mathscr{O}(1)$ multipole moments always induce Planckian curvatures in objects with this compactness.
}
\item[3)]{$\epsilon=\ell^2/m^2\ll1$: this kind of structure is even more compact (we could think about ultracompact wormholes \cite{Visser1989,Visser1989b}, for instance) and, from our perturbative analysis, we can conclude that even tiny multipoles of $M_l\sim\epsilon$ induce Planckian curvatures.
}
\end{itemize}
\noindent
We can understand our results in this paper as a kind of smoothing of the singularities that appear when non-spherical deformations are applied to static event horizons. As shown in Israel's theorem \cite{Israel1967}, these deformations lead to infinities and therefore one must conclude the uniqueness of the Schwarzschild geometry. In fact, we can recover smoothly Israel's theorem if we choose $L$ and $\epsilon$ as independent parameters (so that $X_\epsilon$ becomes a function of them) and take then the limit $\epsilon\rightarrow0$. From Eq. \eqref{eq:mc1}, it is straightforward to show that keeping $L$ fixed in the limit $\epsilon\rightarrow0$ implies that $X_\epsilon$ vanishes as $X_\epsilon\propto\epsilon$. Then, equations that depend on $X_\epsilon$ (and perhaps $\epsilon$) become only dependent on $\epsilon$; an example is given by Eq. \eqref{eq:rhopsiconst}.

In our case, the spacetime curvature remains always bounded due to the fact that $\epsilon\neq0$, but it grows monotonically (for multipole moments with fixed values) as $\epsilon\rightarrow 0$. Even if curvature remains bounded, arbitrarily high curvatures must also be regarded suspiciously. Actually, it is broadly expected that general relativity is an effective theory valid below the Planck curvature (see \cite{Burgess2003} for instance), which would fix $L\propto\ell$ with $\ell$ the Planck length. Let us stress that this does not necessarily implies that one cannot maintain the effective notion of a classical geometry, as it may be the case that an effective field theory that includes in its action higher-order terms in the curvature still provides an adequate description (see, e.g., \cite{BeltranJimenez2017,Heisenberg2018}). A convenient way of rephrasing our main result is then the following: deforming extremely compact objects would quickly push us beyond the regime of applicability of general relativity.

From an observational perspective, the generalized no-hair theorem allows us to devise an independent test of general relativity in vacuum. Our result is translated in terms of constraints between three parameters, that we may take to be the observed spacetime curvature $1/L_{\rm obs}^2$ around the gravitational radius, the innermost redshift that can be probed as parametrized by $\epsilon$ (see \cite{Carballo-Rubio2018b,Carballo-Rubio2018} for a thorough discussion), and the strength of multipoles as measured by $M_l$. We just need to rearrange equation \eqref{eq:mc2} to show that, if general relativity is valid, then
\begin{equation}\label{eq:lastconst}
M_l\lesssim \left(\frac{m}{L_{\rm obs}}\right)^2\epsilon^2.
\end{equation}
Let us imagine for instance what would happen if observations conclude that $L_{\rm obs}\sim m$ and that $\epsilon\ll 1$; $M_l$ should be then proportional to $\epsilon^2$. This could be compared with other independent determinations of these parameters through observations of the environment of astrophysical black holes, such as the ones that will be carried out by the Event Horizon Telescope \cite{Ricarte2014,Psaltis2018}. Hence, Eqs. \eqref{eq:mc2} or \eqref{eq:lastconst} provide a new test of general relativity in vacuum up to the value of the redshift parametrized by $\epsilon$, regardless of the nature of spacetime enclosed by the corresponding equipotential surface (namely, the white region in Fig. \ref{fig:fig1}).

During the last stages of the preparation of this manuscript, we noticed the existence of the recent work \cite{Raposo2018} that deals with the same problem but from a different perspective, and using a different approach (namely, constructing explicitly a family of geometries that include perturbative deviations from the Schwarzschild solution). We wanted to remark that the results of both works are, when overlapping, compatible. Also, the paper \cite{Glampedakis2017} is related to some of the ideas discussed here, focusing its discussion instead on some of the possible implications that a different multipolar structure would have for gravitational waves.

\acknowledgments
Financial support was provided by the Spanish Government through the projects FIS2017-86497-C2-1-P, FIS2017-86497-C2-2-P (with FEDER contribution), FIS2016- 78859-P (AEI/FEDER,UE), and by the Junta de Andalucia through the project FQM219. CB and RCR would like to thank Jos\'e Luis Jaramillo for useful discussions that planted the motivation for the analysis of this problem, in particular, during the X JARRAMPLAS meeting held in April 2018.

\bibliography{refs}

\begin{thebibliography}{33}%
\makeatletter
\providecommand \@ifxundefined [1]{%
 \@ifx{#1\undefined}
}%
\providecommand \@ifnum [1]{%
 \ifnum #1\expandafter \@firstoftwo
 \else \expandafter \@secondoftwo
 \fi
}%
\providecommand \@ifx [1]{%
 \ifx #1\expandafter \@firstoftwo
 \else \expandafter \@secondoftwo
 \fi
}%
\providecommand \natexlab [1]{#1}%
\providecommand \enquote  [1]{``#1''}%
\providecommand \bibnamefont  [1]{#1}%
\providecommand \bibfnamefont [1]{#1}%
\providecommand \citenamefont [1]{#1}%
\providecommand \href@noop [0]{\@secondoftwo}%
\providecommand \href [0]{\begingroup \@sanitize@url \@href}%
\providecommand \@href[1]{\@@startlink{#1}\@@href}%
\providecommand \@@href[1]{\endgroup#1\@@endlink}%
\providecommand \@sanitize@url [0]{\catcode `\\12\catcode `\$12\catcode
  `\&12\catcode `\#12\catcode `\^12\catcode `\_12\catcode `\%12\relax}%
\providecommand \@@startlink[1]{}%
\providecommand \@@endlink[0]{}%
\providecommand \url  [0]{\begingroup\@sanitize@url \@url }%
\providecommand \@url [1]{\endgroup\@href {#1}{\urlprefix }}%
\providecommand \urlprefix  [0]{URL }%
\providecommand \Eprint [0]{\href }%
\providecommand \doibase [0]{http://dx.doi.org/}%
\providecommand \selectlanguage [0]{\@gobble}%
\providecommand \bibinfo  [0]{\@secondoftwo}%
\providecommand \bibfield  [0]{\@secondoftwo}%
\providecommand \translation [1]{[#1]}%
\providecommand \BibitemOpen [0]{}%
\providecommand \bibitemStop [0]{}%
\providecommand \bibitemNoStop [0]{.\EOS\space}%
\providecommand \EOS [0]{\spacefactor3000\relax}%
\providecommand \BibitemShut  [1]{\csname bibitem#1\endcsname}%
\let\auto@bib@innerbib\@empty
%</preamble>
\bibitem [{\citenamefont {Israel}(1967)}]{Israel1967}%
  \BibitemOpen
  \bibfield  {author} {\bibinfo {author} {\bibfnamefont {W.}~\bibnamefont
  {Israel}},\ }\bibfield  {title} {\enquote {\bibinfo {title} {{Event horizons
  in static vacuum space-times}},}\ }\href {\doibase 10.1103/PhysRev.164.1776}
  {\bibfield  {journal} {\bibinfo  {journal} {Phys. Rev.}\ }\textbf {\bibinfo
  {volume} {164}},\ \bibinfo {pages} {1776--1779} (\bibinfo {year}
  {1967})}\BibitemShut {NoStop}%
%%CITATION = PHRVA,164,1776;%%
\bibitem [{\citenamefont {Carter}(1971)}]{Carter1971}%
  \BibitemOpen
  \bibfield  {author} {\bibinfo {author} {\bibfnamefont {B.}~\bibnamefont
  {Carter}},\ }\bibfield  {title} {\enquote {\bibinfo {title} {{Axisymmetric
  Black Hole Has Only Two Degrees of Freedom}},}\ }\href {\doibase
  10.1103/PhysRevLett.26.331} {\bibfield  {journal} {\bibinfo  {journal} {Phys.
  Rev. Lett.}\ }\textbf {\bibinfo {volume} {26}},\ \bibinfo {pages} {331--333}
  (\bibinfo {year} {1971})}\BibitemShut {NoStop}%
%%CITATION = PRLTA,26,331;%%
\bibitem [{\citenamefont {Robinson}(1975)}]{Robinson1975}%
  \BibitemOpen
  \bibfield  {author} {\bibinfo {author} {\bibfnamefont {D.~C.}\ \bibnamefont
  {Robinson}},\ }\bibfield  {title} {\enquote {\bibinfo {title} {{Uniqueness of
  the Kerr black hole}},}\ }\href {\doibase 10.1103/PhysRevLett.34.905}
  {\bibfield  {journal} {\bibinfo  {journal} {Phys. Rev. Lett.}\ }\textbf
  {\bibinfo {volume} {34}},\ \bibinfo {pages} {905--906} (\bibinfo {year}
  {1975})}\BibitemShut {NoStop}%
%%CITATION = PRLTA,34,905;%%
\bibitem [{\citenamefont {Wald}(1984)}]{Wald1984}%
  \BibitemOpen
  \bibfield  {author} {\bibinfo {author} {\bibfnamefont {R.~M.}\ \bibnamefont
  {Wald}},\ }\href {\doibase 10.7208/chicago/9780226870373.001.0001} {\emph
  {\bibinfo {title} {{General Relativity}}}}\ (\bibinfo  {publisher} {Chicago
  Univ. Pr.},\ \bibinfo {address} {Chicago, USA},\ \bibinfo {year}
  {1984})\BibitemShut {NoStop}%
%%CITATION = INSPIRE-209356;%%
\bibitem [{\citenamefont {{Carter}}(1987)}]{Carter1987}%
  \BibitemOpen
  \bibfield  {author} {\bibinfo {author} {\bibfnamefont {B.}~\bibnamefont
  {{Carter}}},\ }\bibfield  {title} {\enquote {\bibinfo {title} {{Mathematical
  foundations of the theory of relativistic stellar and black hole
  configurations.}}}\ }in\ \href@noop {} {\emph {\bibinfo {booktitle}
  {Gravitation in Astrophysics - Carg{\`e}se 1986}}},\ \bibinfo {editor}
  {edited by\ \bibinfo {editor} {\bibfnamefont {B.}~\bibnamefont {{Carter}}}\
  and\ \bibinfo {editor} {\bibfnamefont {J.~B.}\ \bibnamefont {{Hartle}}}}\
  (\bibinfo {year} {1987})\ pp.\ \bibinfo {pages} {63--122}\BibitemShut
  {NoStop}%
\bibitem [{\citenamefont {Carter}(1997)}]{Carter1997}%
  \BibitemOpen
  \bibfield  {author} {\bibinfo {author} {\bibfnamefont {B.}~\bibnamefont
  {Carter}},\ }\bibfield  {title} {\enquote {\bibinfo {title} {{Has the black
  hole equilibrium problem been solved?}}}\ }in\ \href@noop {} {\emph {\bibinfo
  {booktitle} {{Recent developments in theoretical and experimental general
  relativity, gravitation, and relativistic field theories. Proceedings, 8th
  Marcel Grossmann meeting, MG8, Jerusalem, Israel, June 22-27, 1997. Pts. A,
  B}}}}\ (\bibinfo {year} {1997})\ pp.\ \bibinfo {pages} {136--155},\ \Eprint
  {http://arxiv.org/abs/gr-qc/9712038} {arXiv:gr-qc/9712038 [gr-qc]}
  \BibitemShut {NoStop}%
%%CITATION = GR-QC/9712038;%%
\bibitem [{\citenamefont {Mazur}(2000)}]{Mazur2000}%
  \BibitemOpen
  \bibfield  {author} {\bibinfo {author} {\bibfnamefont {P.~O.}\ \bibnamefont
  {Mazur}},\ }\bibfield  {title} {\enquote {\bibinfo {title} {{Black hole
  uniqueness theorems}},}\ }\href@noop {} {\  (\bibinfo {year} {2000})},\
  \Eprint {http://arxiv.org/abs/hep-th/0101012} {arXiv:hep-th/0101012 [hep-th]}
  \BibitemShut {NoStop}%
%%CITATION = HEP-TH/0101012;%%
\bibitem [{\citenamefont {Johannsen}\ and\ \citenamefont
  {Psaltis}(2011)}]{Johannsen2010}%
  \BibitemOpen
  \bibfield  {author} {\bibinfo {author} {\bibfnamefont {T.}~\bibnamefont
  {Johannsen}}\ and\ \bibinfo {author} {\bibfnamefont {D.}~\bibnamefont
  {Psaltis}},\ }\bibfield  {title} {\enquote {\bibinfo {title} {{Testing the
  No-Hair Theorem with Observations of Black Holes in the Electromagnetic
  Spectrum}},}\ }\href {\doibase 10.1016/j.asr.2010.10.019} {\bibfield
  {journal} {\bibinfo  {journal} {Adv. Space Res.}\ }\textbf {\bibinfo {volume}
  {47}},\ \bibinfo {pages} {528--532} (\bibinfo {year} {2011})},\ \Eprint
  {http://arxiv.org/abs/1008.3902} {arXiv:1008.3902 [astro-ph.HE]} \BibitemShut
  {NoStop}%
%%CITATION = ARXIV:1008.3902;%%
\bibitem [{\citenamefont {Bambi}\ \emph {et~al.}(2016)\citenamefont {Bambi},
  \citenamefont {Jiang},\ and\ \citenamefont {Steiner}}]{Bambi2015}%
  \BibitemOpen
  \bibfield  {author} {\bibinfo {author} {\bibfnamefont {C.}~\bibnamefont
  {Bambi}}, \bibinfo {author} {\bibfnamefont {J.}~\bibnamefont {Jiang}}, \ and\
  \bibinfo {author} {\bibfnamefont {J.~F.}\ \bibnamefont {Steiner}},\
  }\bibfield  {title} {\enquote {\bibinfo {title} {{Testing the no-hair theorem
  with the continuum-fitting and the iron line methods: a short review}},}\
  }\href {\doibase 10.1088/0264-9381/33/6/064001} {\bibfield  {journal}
  {\bibinfo  {journal} {Class. Quant. Grav.}\ }\textbf {\bibinfo {volume}
  {33}},\ \bibinfo {pages} {064001} (\bibinfo {year} {2016})},\ \Eprint
  {http://arxiv.org/abs/1511.07587} {arXiv:1511.07587 [gr-qc]} \BibitemShut
  {NoStop}%
%%CITATION = ARXIV:1511.07587;%%
\bibitem [{\citenamefont {Cardoso}\ and\ \citenamefont
  {Gualtieri}(2016)}]{Cardoso2016}%
  \BibitemOpen
  \bibfield  {author} {\bibinfo {author} {\bibfnamefont {V.}~\bibnamefont
  {Cardoso}}\ and\ \bibinfo {author} {\bibfnamefont {L.}~\bibnamefont
  {Gualtieri}},\ }\bibfield  {title} {\enquote {\bibinfo {title} {{Testing the
  black hole ``no-hair" hypothesis}},}\ }\href {\doibase
  10.1088/0264-9381/33/17/174001} {\bibfield  {journal} {\bibinfo  {journal}
  {Class. Quant. Grav.}\ }\textbf {\bibinfo {volume} {33}},\ \bibinfo {pages}
  {174001} (\bibinfo {year} {2016})},\ \Eprint
  {http://arxiv.org/abs/1607.03133} {arXiv:1607.03133 [gr-qc]} \BibitemShut
  {NoStop}%
%%CITATION = ARXIV:1607.03133;%%
\bibitem [{\citenamefont {Krishnendu}\ \emph {et~al.}(2017)\citenamefont
  {Krishnendu}, \citenamefont {Arun},\ and\ \citenamefont
  {Mishra}}]{Krishnendu2017}%
  \BibitemOpen
  \bibfield  {author} {\bibinfo {author} {\bibfnamefont {N.~V.}\ \bibnamefont
  {Krishnendu}}, \bibinfo {author} {\bibfnamefont {K.~G.}\ \bibnamefont
  {Arun}}, \ and\ \bibinfo {author} {\bibfnamefont {C.~K.}\ \bibnamefont
  {Mishra}},\ }\bibfield  {title} {\enquote {\bibinfo {title} {{Testing the
  binary black hole nature of a compact binary coalescence}},}\ }\href
  {\doibase 10.1103/PhysRevLett.119.091101} {\bibfield  {journal} {\bibinfo
  {journal} {Phys. Rev. Lett.}\ }\textbf {\bibinfo {volume} {119}},\ \bibinfo
  {pages} {091101} (\bibinfo {year} {2017})},\ \Eprint
  {http://arxiv.org/abs/1701.06318} {arXiv:1701.06318 [gr-qc]} \BibitemShut
  {NoStop}%
%%CITATION = ARXIV:1701.06318;%%
\bibitem [{\citenamefont {Thrane}\ \emph {et~al.}(2017)\citenamefont {Thrane},
  \citenamefont {Lasky},\ and\ \citenamefont {Levin}}]{Thrane2017}%
  \BibitemOpen
  \bibfield  {author} {\bibinfo {author} {\bibfnamefont {E.}~\bibnamefont
  {Thrane}}, \bibinfo {author} {\bibfnamefont {P.~D.}\ \bibnamefont {Lasky}}, \
  and\ \bibinfo {author} {\bibfnamefont {Y.}~\bibnamefont {Levin}},\ }\bibfield
   {title} {\enquote {\bibinfo {title} {{Challenges testing the no-hair theorem
  with gravitational waves}},}\ }\href {\doibase 10.1103/PhysRevD.96.102004}
  {\bibfield  {journal} {\bibinfo  {journal} {Phys. Rev.}\ }\textbf {\bibinfo
  {volume} {D96}},\ \bibinfo {pages} {102004} (\bibinfo {year} {2017})},\
  \Eprint {http://arxiv.org/abs/1706.05152} {arXiv:1706.05152 [gr-qc]}
  \BibitemShut {NoStop}%
%%CITATION = ARXIV:1706.05152;%%
\bibitem [{\citenamefont {Cardoso}\ and\ \citenamefont
  {Pani}(2017)}]{Cardoso2017}%
  \BibitemOpen
  \bibfield  {author} {\bibinfo {author} {\bibfnamefont {V.}~\bibnamefont
  {Cardoso}}\ and\ \bibinfo {author} {\bibfnamefont {P.}~\bibnamefont {Pani}},\
  }\bibfield  {title} {\enquote {\bibinfo {title} {{Tests for the existence of
  black holes through gravitational wave echoes}},}\ }\href {\doibase
  10.1038/s41550-017-0225-y} {\bibfield  {journal} {\bibinfo  {journal} {Nat.
  Astron.}\ }\textbf {\bibinfo {volume} {1}},\ \bibinfo {pages} {586--591}
  (\bibinfo {year} {2017})},\ \Eprint {http://arxiv.org/abs/1709.01525}
  {arXiv:1709.01525 [gr-qc]} \BibitemShut {NoStop}%
%%CITATION = ARXIV:1709.01525;%%
\bibitem [{\citenamefont {Carballo-Rubio}\ \emph
  {et~al.}(2018{\natexlab{a}})\citenamefont {Carballo-Rubio}, \citenamefont
  {Di~Filippo}, \citenamefont {Liberati},\ and\ \citenamefont
  {Visser}}]{Carballo-Rubio2018b}%
  \BibitemOpen
  \bibfield  {author} {\bibinfo {author} {\bibfnamefont {R.}~\bibnamefont
  {Carballo-Rubio}}, \bibinfo {author} {\bibfnamefont {F.}~\bibnamefont
  {Di~Filippo}}, \bibinfo {author} {\bibfnamefont {S.}~\bibnamefont
  {Liberati}}, \ and\ \bibinfo {author} {\bibfnamefont {M.}~\bibnamefont
  {Visser}},\ }\bibfield  {title} {\enquote {\bibinfo {title}
  {{Phenomenological aspects of black holes beyond general relativity}},}\
  }\href {\doibase 10.1103/PhysRevD.98.124009} {\bibfield  {journal} {\bibinfo
  {journal} {Phys. Rev.}\ }\textbf {\bibinfo {volume} {D98}},\ \bibinfo {pages}
  {124009} (\bibinfo {year} {2018}{\natexlab{a}})},\ \Eprint
  {http://arxiv.org/abs/1809.08238} {arXiv:1809.08238 [gr-qc]} \BibitemShut
  {NoStop}%
%%CITATION = ARXIV:1809.08238;%%
\bibitem [{\citenamefont {Giddings}\ and\ \citenamefont
  {Psaltis}(2018)}]{Giddings2016}%
  \BibitemOpen
  \bibfield  {author} {\bibinfo {author} {\bibfnamefont {S.~B.}\ \bibnamefont
  {Giddings}}\ and\ \bibinfo {author} {\bibfnamefont {D.}~\bibnamefont
  {Psaltis}},\ }\bibfield  {title} {\enquote {\bibinfo {title} {{Event Horizon
  Telescope Observations as Probes for Quantum Structure of Astrophysical Black
  Holes}},}\ }\href {\doibase 10.1103/PhysRevD.97.084035} {\bibfield  {journal}
  {\bibinfo  {journal} {Phys. Rev.}\ }\textbf {\bibinfo {volume} {D97}},\
  \bibinfo {pages} {084035} (\bibinfo {year} {2018})},\ \Eprint
  {http://arxiv.org/abs/1606.07814} {arXiv:1606.07814 [astro-ph.HE]}
  \BibitemShut {NoStop}%
%%CITATION = ARXIV:1606.07814;%%
\bibitem [{\citenamefont {Israel}(1968)}]{Israel1967b}%
  \BibitemOpen
  \bibfield  {author} {\bibinfo {author} {\bibfnamefont {W.}~\bibnamefont
  {Israel}},\ }\bibfield  {title} {\enquote {\bibinfo {title} {{Event horizons
  in static electrovac space-times}},}\ }\href {\doibase 10.1007/BF01645859}
  {\bibfield  {journal} {\bibinfo  {journal} {Commun. Math. Phys.}\ }\textbf
  {\bibinfo {volume} {8}},\ \bibinfo {pages} {245--260} (\bibinfo {year}
  {1968})}\BibitemShut {NoStop}%
%%CITATION = CMPHA,8,245;%%
\bibitem [{\citenamefont {Olver}\ \emph {et~al.}(2010)\citenamefont {Olver},
  \citenamefont {of~Standards}, \citenamefont {(U.S.)}, \citenamefont {Lozier},
  \citenamefont {Boisvert},\ and\ \citenamefont {Clark}}]{Olver2010}%
  \BibitemOpen
  \bibfield  {author} {\bibinfo {author} {\bibfnamefont {F.~W.~J.}\
  \bibnamefont {Olver}}, \bibinfo {author} {\bibfnamefont {National~Institute}\
  \bibnamefont {of~Standards}}, \bibinfo {author} {\bibfnamefont {Technology}\
  \bibnamefont {(U.S.)}}, \bibinfo {author} {\bibfnamefont {D.~W.}\
  \bibnamefont {Lozier}}, \bibinfo {author} {\bibfnamefont {R.~F.}\
  \bibnamefont {Boisvert}}, \ and\ \bibinfo {author} {\bibfnamefont {C.~W.}\
  \bibnamefont {Clark}},\ }\href
  {https://books.google.it/books?id=7EZ6PwDy4qEC} {\emph {\bibinfo {title}
  {NIST Handbook of Mathematical Functions Paperback and CD-ROM}}}\ (\bibinfo
  {publisher} {Cambridge University Press},\ \bibinfo {year}
  {2010})\BibitemShut {NoStop}%
\bibitem [{\citenamefont {Mazur}\ and\ \citenamefont
  {Mottola}(2004)}]{Mazur2004}%
  \BibitemOpen
  \bibfield  {author} {\bibinfo {author} {\bibfnamefont {P.~O.}\ \bibnamefont
  {Mazur}}\ and\ \bibinfo {author} {\bibfnamefont {E.}~\bibnamefont
  {Mottola}},\ }\bibfield  {title} {\enquote {\bibinfo {title} {{Gravitational
  vacuum condensate stars}},}\ }\href {\doibase 10.1073/pnas.0402717101}
  {\bibfield  {journal} {\bibinfo  {journal} {Proc. Nat. Acad. Sci.}\ }\textbf
  {\bibinfo {volume} {101}},\ \bibinfo {pages} {9545--9550} (\bibinfo {year}
  {2004})},\ \Eprint {http://arxiv.org/abs/gr-qc/0407075} {arXiv:gr-qc/0407075
  [gr-qc]} \BibitemShut {NoStop}%
%%CITATION = GR-QC/0407075;%%
\bibitem [{\citenamefont {Mottola}(2010)}]{Mottola2010}%
  \BibitemOpen
  \bibfield  {author} {\bibinfo {author} {\bibfnamefont {E.}~\bibnamefont
  {Mottola}},\ }\bibfield  {title} {\enquote {\bibinfo {title} {{New Horizons
  in Gravity: The Trace Anomaly, Dark Energy and Condensate Stars}},}\
  }\bibfield  {booktitle} {\emph {\bibinfo {booktitle} {{Non-perturbative
  gravity and quantum chromodynamics. Proceedings, 49th Cracow School of
  Theoretical Physics, Zakopane, Poland, May 31-June 10, 2009}}},\ }\href@noop
  {} {\bibfield  {journal} {\bibinfo  {journal} {Acta Phys. Polon.}\ }\textbf
  {\bibinfo {volume} {B41}},\ \bibinfo {pages} {2031--2162} (\bibinfo {year}
  {2010})},\ \Eprint {http://arxiv.org/abs/1008.5006} {arXiv:1008.5006 [gr-qc]}
  \BibitemShut {NoStop}%
%%CITATION = ARXIV:1008.5006;%%
\bibitem [{\citenamefont {Visser}\ and\ \citenamefont
  {Wiltshire}(2004)}]{Visser2003}%
  \BibitemOpen
  \bibfield  {author} {\bibinfo {author} {\bibfnamefont {M.}~\bibnamefont
  {Visser}}\ and\ \bibinfo {author} {\bibfnamefont {D.~L.}\ \bibnamefont
  {Wiltshire}},\ }\bibfield  {title} {\enquote {\bibinfo {title} {{Stable
  gravastars: An Alternative to black holes?}}}\ }\href {\doibase
  10.1088/0264-9381/21/4/027} {\bibfield  {journal} {\bibinfo  {journal}
  {Class. Quant. Grav.}\ }\textbf {\bibinfo {volume} {21}},\ \bibinfo {pages}
  {1135--1152} (\bibinfo {year} {2004})},\ \Eprint
  {http://arxiv.org/abs/gr-qc/0310107} {arXiv:gr-qc/0310107 [gr-qc]}
  \BibitemShut {NoStop}%
%%CITATION = GR-QC/0310107;%%
\bibitem [{\citenamefont {Barcelo}\ \emph {et~al.}(2008)\citenamefont
  {Barcelo}, \citenamefont {Liberati}, \citenamefont {Sonego},\ and\
  \citenamefont {Visser}}]{Barcelo2007}%
  \BibitemOpen
  \bibfield  {author} {\bibinfo {author} {\bibfnamefont {C.}~\bibnamefont
  {Barcelo}}, \bibinfo {author} {\bibfnamefont {S.}~\bibnamefont {Liberati}},
  \bibinfo {author} {\bibfnamefont {S.}~\bibnamefont {Sonego}}, \ and\ \bibinfo
  {author} {\bibfnamefont {M.}~\bibnamefont {Visser}},\ }\bibfield  {title}
  {\enquote {\bibinfo {title} {{Fate of gravitational collapse in semiclassical
  gravity}},}\ }\href {\doibase 10.1103/PhysRevD.77.044032} {\bibfield
  {journal} {\bibinfo  {journal} {Phys. Rev.}\ }\textbf {\bibinfo {volume}
  {D77}},\ \bibinfo {pages} {044032} (\bibinfo {year} {2008})},\ \Eprint
  {http://arxiv.org/abs/0712.1130} {arXiv:0712.1130 [gr-qc]} \BibitemShut
  {NoStop}%
%%CITATION = ARXIV:0712.1130;%%
\bibitem [{\citenamefont {Barcel\'o}\ \emph {et~al.}(2009)\citenamefont
  {Barcel\'o}, \citenamefont {Liberati}, \citenamefont {Sonego},\ and\
  \citenamefont {Visser}}]{Barcelo2009}%
  \BibitemOpen
  \bibfield  {author} {\bibinfo {author} {\bibfnamefont {C.}~\bibnamefont
  {Barcel\'o}}, \bibinfo {author} {\bibfnamefont {S.}~\bibnamefont {Liberati}},
  \bibinfo {author} {\bibfnamefont {S.}~\bibnamefont {Sonego}}, \ and\ \bibinfo
  {author} {\bibfnamefont {M.}~\bibnamefont {Visser}},\ }\bibfield  {title}
  {\enquote {\bibinfo {title} {{Black Stars, Not Holes}},}\ }\href {\doibase
  10.1038/scientificamerican1009-38} {\bibfield  {journal} {\bibinfo  {journal}
  {Sci. Am.}\ }\textbf {\bibinfo {volume} {301}},\ \bibinfo {pages} {38--45}
  (\bibinfo {year} {2009})}\BibitemShut {NoStop}%
%%CITATION = SCAMA,301,38;%%
\bibitem [{\citenamefont {Carballo-Rubio}(2018)}]{Carballo-Rubio2017}%
  \BibitemOpen
  \bibfield  {author} {\bibinfo {author} {\bibfnamefont {R.}~\bibnamefont
  {Carballo-Rubio}},\ }\bibfield  {title} {\enquote {\bibinfo {title} {{Stellar
  equilibrium in semiclassical gravity}},}\ }\href {\doibase
  10.1103/PhysRevLett.120.061102} {\bibfield  {journal} {\bibinfo  {journal}
  {Phys. Rev. Lett.}\ }\textbf {\bibinfo {volume} {120}},\ \bibinfo {pages}
  {061102} (\bibinfo {year} {2018})},\ \Eprint
  {http://arxiv.org/abs/1706.05379} {arXiv:1706.05379 [gr-qc]} \BibitemShut
  {NoStop}%
%%CITATION = ARXIV:1706.05379;%%
\bibitem [{\citenamefont {Visser}(1989{\natexlab{a}})}]{Visser1989}%
  \BibitemOpen
  \bibfield  {author} {\bibinfo {author} {\bibfnamefont {M.}~\bibnamefont
  {Visser}},\ }\bibfield  {title} {\enquote {\bibinfo {title} {{Traversable
  wormholes: Some simple examples}},}\ }\href {\doibase
  10.1103/PhysRevD.39.3182} {\bibfield  {journal} {\bibinfo  {journal} {Phys.
  Rev.}\ }\textbf {\bibinfo {volume} {D39}},\ \bibinfo {pages} {3182--3184}
  (\bibinfo {year} {1989}{\natexlab{a}})},\ \Eprint
  {http://arxiv.org/abs/0809.0907} {arXiv:0809.0907 [gr-qc]} \BibitemShut
  {NoStop}%
%%CITATION = ARXIV:0809.0907;%%
\bibitem [{\citenamefont {Visser}(1989{\natexlab{b}})}]{Visser1989b}%
  \BibitemOpen
  \bibfield  {author} {\bibinfo {author} {\bibfnamefont {M.}~\bibnamefont
  {Visser}},\ }\bibfield  {title} {\enquote {\bibinfo {title} {{Traversable
  wormholes from surgically modified Schwarzschild space-times}},}\ }\href
  {\doibase 10.1016/0550-3213(89)90100-4} {\bibfield  {journal} {\bibinfo
  {journal} {Nucl. Phys.}\ }\textbf {\bibinfo {volume} {B328}},\ \bibinfo
  {pages} {203--212} (\bibinfo {year} {1989}{\natexlab{b}})},\ \Eprint
  {http://arxiv.org/abs/0809.0927} {arXiv:0809.0927 [gr-qc]} \BibitemShut
  {NoStop}%
%%CITATION = ARXIV:0809.0927;%%
\bibitem [{\citenamefont {Burgess}(2004)}]{Burgess2003}%
  \BibitemOpen
  \bibfield  {author} {\bibinfo {author} {\bibfnamefont {C.~P.}\ \bibnamefont
  {Burgess}},\ }\bibfield  {title} {\enquote {\bibinfo {title} {{Quantum
  gravity in everyday life: General relativity as an effective field
  theory}},}\ }\href {\doibase 10.12942/lrr-2004-5} {\bibfield  {journal}
  {\bibinfo  {journal} {Living Rev. Rel.}\ }\textbf {\bibinfo {volume} {7}},\
  \bibinfo {pages} {5--56} (\bibinfo {year} {2004})},\ \Eprint
  {http://arxiv.org/abs/gr-qc/0311082} {arXiv:gr-qc/0311082 [gr-qc]}
  \BibitemShut {NoStop}%
%%CITATION = GR-QC/0311082;%%
\bibitem [{\citenamefont {Beltran~Jimenez}\ \emph {et~al.}(2018)\citenamefont
  {Beltran~Jimenez}, \citenamefont {Heisenberg}, \citenamefont {Olmo},\ and\
  \citenamefont {Rubiera-Garcia}}]{BeltranJimenez2017}%
  \BibitemOpen
  \bibfield  {author} {\bibinfo {author} {\bibfnamefont {J.}~\bibnamefont
  {Beltran~Jimenez}}, \bibinfo {author} {\bibfnamefont {L.}~\bibnamefont
  {Heisenberg}}, \bibinfo {author} {\bibfnamefont {G.~J.}\ \bibnamefont
  {Olmo}}, \ and\ \bibinfo {author} {\bibfnamefont {D.}~\bibnamefont
  {Rubiera-Garcia}},\ }\bibfield  {title} {\enquote {\bibinfo {title}
  {{Born-Infeld inspired modifications of gravity}},}\ }\href {\doibase
  10.1016/j.physrep.2017.11.001} {\bibfield  {journal} {\bibinfo  {journal}
  {Phys. Rept.}\ }\textbf {\bibinfo {volume} {727}},\ \bibinfo {pages} {1--129}
  (\bibinfo {year} {2018})},\ \Eprint {http://arxiv.org/abs/1704.03351}
  {arXiv:1704.03351 [gr-qc]} \BibitemShut {NoStop}%
%%CITATION = ARXIV:1704.03351;%%
\bibitem [{\citenamefont {Heisenberg}(2019)}]{Heisenberg2018}%
  \BibitemOpen
  \bibfield  {author} {\bibinfo {author} {\bibfnamefont {Lavinia}\ \bibnamefont
  {Heisenberg}},\ }\bibfield  {title} {\enquote {\bibinfo {title} {{A
  systematic approach to generalisations of General Relativity and their
  cosmological implications}},}\ }\href {\doibase
  10.1016/j.physrep.2018.11.006} {\bibfield  {journal} {\bibinfo  {journal}
  {Phys. Rept.}\ }\textbf {\bibinfo {volume} {796}},\ \bibinfo {pages} {1--113}
  (\bibinfo {year} {2019})},\ \Eprint {http://arxiv.org/abs/1807.01725}
  {arXiv:1807.01725 [gr-qc]} \BibitemShut {NoStop}%
%%CITATION = ARXIV:1807.01725;%%
\bibitem [{\citenamefont {Carballo-Rubio}\ \emph
  {et~al.}(2018{\natexlab{b}})\citenamefont {Carballo-Rubio}, \citenamefont
  {Kumar},\ and\ \citenamefont {Lu}}]{Carballo-Rubio2018}%
  \BibitemOpen
  \bibfield  {author} {\bibinfo {author} {\bibfnamefont {R.}~\bibnamefont
  {Carballo-Rubio}}, \bibinfo {author} {\bibfnamefont {P.}~\bibnamefont
  {Kumar}}, \ and\ \bibinfo {author} {\bibfnamefont {W.}~\bibnamefont {Lu}},\
  }\bibfield  {title} {\enquote {\bibinfo {title} {{Seeking observational
  evidence for the formation of trapping horizons in astrophysical black
  holes}},}\ }\href {\doibase 10.1103/PhysRevD.97.123012} {\bibfield  {journal}
  {\bibinfo  {journal} {Phys. Rev.}\ }\textbf {\bibinfo {volume} {D97}},\
  \bibinfo {pages} {123012} (\bibinfo {year} {2018}{\natexlab{b}})},\ \Eprint
  {http://arxiv.org/abs/1804.00663} {arXiv:1804.00663 [gr-qc]} \BibitemShut
  {NoStop}%
%%CITATION = ARXIV:1804.00663;%%
\bibitem [{\citenamefont {Ricarte}\ and\ \citenamefont
  {Dexter}(2015)}]{Ricarte2014}%
  \BibitemOpen
  \bibfield  {author} {\bibinfo {author} {\bibfnamefont {A.}~\bibnamefont
  {Ricarte}}\ and\ \bibinfo {author} {\bibfnamefont {J.}~\bibnamefont
  {Dexter}},\ }\bibfield  {title} {\enquote {\bibinfo {title} {{The Event
  Horizon Telescope: exploring strong gravity and accretion physics}},}\ }\href
  {\doibase 10.1093/mnras/stu2128} {\bibfield  {journal} {\bibinfo  {journal}
  {Mon. Not. Roy. Astron. Soc.}\ }\textbf {\bibinfo {volume} {446}},\ \bibinfo
  {pages} {1973--1987} (\bibinfo {year} {2015})},\ \Eprint
  {http://arxiv.org/abs/1410.2899} {arXiv:1410.2899 [astro-ph.HE]} \BibitemShut
  {NoStop}%
%%CITATION = ARXIV:1410.2899;%%
\bibitem [{\citenamefont {Psaltis}(2018)}]{Psaltis2018}%
  \BibitemOpen
  \bibfield  {author} {\bibinfo {author} {\bibfnamefont {D.}~\bibnamefont
  {Psaltis}},\ }\bibfield  {title} {\enquote {\bibinfo {title} {{Testing
  General Relativity with the Event Horizon Telescope}},}\ }\href@noop {} {\
  (\bibinfo {year} {2018})},\ \Eprint {http://arxiv.org/abs/1806.09740}
  {arXiv:1806.09740 [astro-ph.HE]} \BibitemShut {NoStop}%
%%CITATION = ARXIV:1806.09740;%%
\bibitem [{\citenamefont {Raposo}\ \emph {et~al.}(2019)\citenamefont {Raposo},
  \citenamefont {Pani},\ and\ \citenamefont {Emparan}}]{Raposo2018}%
  \BibitemOpen
  \bibfield  {author} {\bibinfo {author} {\bibfnamefont {Guilherme}\
  \bibnamefont {Raposo}}, \bibinfo {author} {\bibfnamefont {Paolo}\
  \bibnamefont {Pani}}, \ and\ \bibinfo {author} {\bibfnamefont {Roberto}\
  \bibnamefont {Emparan}},\ }\bibfield  {title} {\enquote {\bibinfo {title}
  {{Exotic compact objects with soft hair}},}\ }\href {\doibase
  10.1103/PhysRevD.99.104050} {\bibfield  {journal} {\bibinfo  {journal} {Phys.
  Rev.}\ }\textbf {\bibinfo {volume} {D99}},\ \bibinfo {pages} {104050}
  (\bibinfo {year} {2019})},\ \Eprint {http://arxiv.org/abs/1812.07615}
  {arXiv:1812.07615 [gr-qc]} \BibitemShut {NoStop}%
%%CITATION = ARXIV:1812.07615;%%
\bibitem [{\citenamefont {Glampedakis}\ and\ \citenamefont
  {Pappas}(2018)}]{Glampedakis2017}%
  \BibitemOpen
  \bibfield  {author} {\bibinfo {author} {\bibfnamefont {K.}~\bibnamefont
  {Glampedakis}}\ and\ \bibinfo {author} {\bibfnamefont {G.}~\bibnamefont
  {Pappas}},\ }\bibfield  {title} {\enquote {\bibinfo {title} {{How well can
  ultracompact bodies imitate black hole ringdowns?}}}\ }\href {\doibase
  10.1103/PhysRevD.97.041502} {\bibfield  {journal} {\bibinfo  {journal} {Phys.
  Rev.}\ }\textbf {\bibinfo {volume} {D97}},\ \bibinfo {pages} {041502}
  (\bibinfo {year} {2018})},\ \Eprint {http://arxiv.org/abs/1710.02136}
  {arXiv:1710.02136 [gr-qc]} \BibitemShut {NoStop}%
%%CITATION = ARXIV:1710.02136;%%
\end{thebibliography}%

\end{document}